\documentclass[aps,prfluids,onecolumn,superscriptaddress,amsmath]{revtex4-2}
\usepackage{bm}
\usepackage{layouts}
\usepackage{graphicx}
\usepackage[hidelinks=True]{hyperref}
\usepackage{siunitx}

\newcommand{\grad}{\bm{\nabla}}
\newcommand{\bu}{\bm{u}}
\newcommand{\pd}{\partial}

\begin{document}


\title{Double-diffusive transport in multicomponent vertical convection}


\author{Christopher J. Howland}
\email{c.j.howland@utwente.nl}
\affiliation{Physics of Fluids Group, Max Planck Center for Complex Fluid Dynamics, and J.M. Burgers Centre for Fluid Dynamics, University of Twente, P.O. Box 217, 7500AE Enschede, Netherlands}
\author{Roberto Verzicco}
\affiliation{Dipartimento di Ingegneria Industriale, University of Rome `Tor Vergata', Via del Politecnico 1, Roma 00133, Italy}
\affiliation{Gran Sasso Science Institute, Viale F. Crispi, 7, 67100 L'Aquila, Italy}
\affiliation{Physics of Fluids Group, Max Planck Center for Complex Fluid Dynamics, and J.M. Burgers Centre for Fluid Dynamics, University of Twente, P.O. Box 217, 7500AE Enschede, Netherlands}
\author{Detlef Lohse}
\email{d.lohse@utwente.nl}
\affiliation{Physics of Fluids Group, Max Planck Center for Complex Fluid Dynamics, and J.M. Burgers Centre for Fluid Dynamics, University of Twente, P.O. Box 217, 7500AE Enschede, Netherlands}
\affiliation{Max Planck Institute for Dynamics and Self-Organization, Am Fassberg 17, 37077 G\"{o}ttingen, Germany}


\date{\today}

\begin{abstract}
    Motivated by the ablation of vertical ice faces in salt water, we use three-dimensional direct numerical simulations to investigate the heat and salt fluxes in two-scalar vertical convection.
    For parameters relevant to ice-ocean interfaces in the convection-dominated regime, we observe that the salinity field drives the convection and that heat is essentially transported as a passive scalar.
    By varying the diffusivity ratio of heat and salt (i.e., the Lewis number $Le$), we identify how the different molecular diffusivities affect the scalar fluxes through the system.
    Away from the walls, we find that the heat transport is determined by a turbulent Prandtl number of $Pr_t\approx 1$ and that double-diffusive effects are practically negligible.
    However, the difference in molecular diffusivities plays an important role close to the boundaries.
    In the (unrealistic) case where salt diffused faster than heat, the ratio of salt-to-heat fluxes would scale as $Le^{1/3}$, consistent with classical nested scalar boundary layers.
    However, in the realistic case of faster heat diffusion (relative to salt), we observe a transition towards a $Le^{1/2}$ scaling of the ratio of the fluxes.
    This coincides with the thermal boundary layer width growing beyond the thickness of the viscous boundary layer.
    We find that this transition
    is not determined by a critical Lewis number, but rather by a critical Prandtl number $Pr\approx 10$, slightly below that for cold seawater where $Pr=14$.
    We compare our results to similar studies of sheared and double-diffusive flow under ice shelves, and discuss the implications for fluxes in large-scale ice-ocean models.
    By coupling our results to ice-ocean interface thermodynamics, we describe how the flux ratio impacts the interfacial salinity, and hence the strength of solutal convection and the ablation rate.
\end{abstract}


\maketitle

\section{Introduction \label{sec:intro}}

Over the last century, the loss of land-based ice from the Greenland and Arctic ice sheets has contributed significantly to sea level rise, and the rate of this mass loss has increased up to sixfold over the last 40 years \citep{mouginot_forty-six_2019,rignot_four_2019}.
Future projections from an ensemble of climate models indicate that this rate is set to increase further over the coming century for a range of emissions scenarios, endangering many regions to coastal flooding \citep{goelzer_future_2020,seroussi_ismip6_2020,edwards_projected_2021}.
Despite the importance of these projections, the complexity of the climate system introduces significant uncertainty regarding the magnitude of future sea level rise.
One key source of uncertainty arises from the parameterisation of melting at the ice-ocean interface \citep{favier_assessment_2019}.
These parameterisations range in complexity from simple linear or quadratic dependences on the ambient ocean temperature to buoyant plume models.
To reduce the uncertainty associated with such parameterisations, it is important to understand the physical mechanisms driving the ice ablation, particularly for regions of relatively warm, salty water where ice retreat is fastest.

From a fundamental physical perspective, the melt rate of ice in salt water depends only on the gradients of temperature and concentration at the ice interface, which determine the diffusive fluxes of heat and salt towards the ice \citep{martin_experimental_1977,malyarenko_synthesis_2020}.
A common assumption in melt parameterisations is that the fluxes are determined by the velocity of the water adjacent to the ice, with the flow taking the form of a classical shear-driven turbulent boundary layer \citep{holland_modeling_1999}.
In that case, both scalars (heat and salt) are transported passively.
However, recent observational and experimental work points to buoyancy playing an important role in scenarios where the ambient currents are weak.
Close to the ice-ocean interface, buoyancy perturbations are dominated by differences in the salt concentration of the water rather than temperature.
For horizontal ice faces, this creates a stable density stratification as the cold, fresh meltwater remains in contact with the ice.
This fresh layer can then undergo double-diffusive convection due to the differing diffusivities of heat and salt \citep{kimura_estimation_2015}, leading to observations where the melt rate is independent of the ambient turbulence \citep{middleton_double_2022}.
At steeply sloped ice faces, found at tidewater glaciers \citep{jackson_meltwater_2020} and on the underside of ice shelves (where step-like terraces can form in the basal topography) \citep{dutrieux_basal_2014}, the fresh meltwater instead forms a rising plume \citep{hewitt_subglacial_2020}.
Experiments suggest that the melt rate in this case is also independent of the flow velocity, and that theory for vertical surfaces can easily be applied to those with steep slopes \citep{kerr_dissolution_2015,mcconnochie_dissolution_2018}.

One extreme difficulty for modelling the melt rate of ice in salt water is that the diffusive boundary layers controlling the heat and salt fluxes are on the millimetre scale.
These boundary layers are extremely difficult to analyse experimentally or in the field, but have recently become accessible through numerical simulations.
Resolving the boundary layers allows us to directly measure the diffusive fluxes at the ice interface, which are not only coupled to the melt rate but also to the local melting temperature of the ice, which depends on the local salinity \citep{malyarenko_synthesis_2020}.
Two recent studies have found that stable buoyancy gradients modify the ratio of salt flux to heat flux at the interface when compared to purely shear-driven systems \citep{vreugdenhil_stratification_2019,rosevear_role_2021}.
Through the coupled boundary condition at the ice-water interface, this ratio in turn modifies the melt rate.

The thermodynamic boundary conditions at an ice-ocean interface consist of the liquidus condition (describing how the melt temperature $T_i$ depends on the interfacial salt concentration $C_i$), along with conservation of heat and conservation of salt:
\begin{align}
    T_i + \lambda C_i &= 0, &
    \frac{L}{c_p} \mathcal{V} &= F_T, &
    C_i \mathcal{V} &= F_C . \label{eq:ice_BC}
\end{align}
Here, $\lambda$ is the liquidus slope, $L$ is latent heat, $c_p$ is specific heat capacity, and $\mathcal{V}$ is the ablation velocity of the interface.
We have neglected heat and salt fluxes through the solid ice, such that the interface evolution is purely forced by the diffusive fluxes of heat and salt $F_T$ and $F_C$ from the liquid.
Since the diffusive boundary layers are so small at ice-ocean interfaces, these fluxes need parameterisation in larger-scale models.
Such parameterisations typically arise from theory describing the dimensionless fluxes or Nusselt numbers $Nu$, so when considering the flux ratio from a theoretical perspective, it makes sense to consider a ratio of Nusselt numbers
\begin{equation}
    R = \frac{Nu_C}{Nu_T} = \frac{F_C}{\kappa_C \Delta C/H} \frac{\kappa_T \Delta T/H}{F_T} .
\end{equation}
Eliminating $\mathcal{V}$ from the last two equations of \eqref{eq:ice_BC} then shows us how the flux ratio $R$ determines the interface salinity $C_i$:
\begin{equation}
    \frac{C_i}{\Delta C} = \frac{\mathcal{S}}{Le} R , \label{eq:Ci_R}
\end{equation}
where the ratio of molecular diffusivities $Le=\kappa_T/\kappa_C$ is the Lewis number, and $\mathcal{S}=L/(c_p\Delta T)$ is the Stefan number.
Although \eqref{eq:Ci_R} appears simple, nonlinearity is hidden in $\Delta C = C - C_i$ and $\Delta T$ which depends on $C_i$ through the liquidus condition.
Nevertheless, given a prescribed far-field temperature and concentration value, $R$ uniquely determines the interface concentration through \eqref{eq:Ci_R}.
We elaborate on this point later in \S\ref{sec:discussion}.


The physical mechanisms underlying the aforementioned changes in the flux ratio $R$ are however complex.
In the case of a horizontal ice surface, simulations of diffusive convection beneath a melting ice face \citep{keitzl_reconciling_2016} have found a non-trivial dependence of the flux ratio on the Lewis number $Le$.
Although $Le$ is a fixed value in reality, determined by the fluid properties, realistic values are notoriously difficult to simulate numerically.
In \citep{keitzl_reconciling_2016} and in the current study, the $Le$-dependence of the flux ratio $R$ is investigated to determine the physical mechanisms underlying the value of the flux ratio.
For vertical ice faces, the appropriate flux ratio is completely unknown, with proposed theory \citep{kerr_dissolution_2015} and common parameterisations \citep{jenkins_convection-driven_2011} in disagreement.
Such parameterisations as in \citep{jenkins_convection-driven_2011} are often directly applied as a boundary condition in large-scale modelling studies of plumes at ice-ocean interfaces \citep{xu_subaqueous_2013,sciascia_seasonal_2013,kimura_effect_2014}, so understanding the physics at the boundary is vital for accurate estimates of melt rate and freshwater production.

To gain physical insight into the mechanisms determining the flux ratio in such convective boundary layers, in this paper we perform direct numerical simulations of a highly simplified setup.
We consider the vertical convection (VC) flow \citep{ng_vertical_2015,shishkina_momentum_2016} in an infinite vertical channel between two stationary walls held at fixed (but different) temperatures and salt concentrations.
The fixed scalar values are justified by the results of \citep{gayen_simulation_2016}, where the interfacial values of temperature and salinity at a melting ice face reach a constant value as the flow develops a statistically steady state.
Obviously, at a real ice face in the ocean, local interface temperatures and salinities vary according to the local fluxes, but it is common in the ice-ocean modelling literature to assume that these small-scale fluctuations do not have a significant impact on the adjacent fluid flow \citep{holland_modeling_1999,wells_geophysical-scale_2008}.
Rather than fixing the fluid properties to realistic values, in order to better understand the physical mechanisms, we vary the Schmidt number and Lewis number and systematically investigate how the fluxes depend on the dimensionless control parameters of the system.
This study builds on our previous work on VC at high Prandtl number \citep{howland_boundary_2022}.
As in that study, we use a multiple-resolution technique to perform large three-dimensional simulations with low-diffusivity scalars at a reduced computational cost.
Unlike some previous studies of multicomponent convection in a vertical channel \citep{kerr_double-diffusive_1999}, we neglect the effect of any mean ambient stratification.
In the motivating example of convection at a tidewater glacier face, the buoyancy perturbations in the boundary layer are significantly greater than the buoyancy differences in the ambient, so we do not expect detrainment from the wall and layering due to stratification, at least at the scales we are considering.
On larger scales, the entrainment of salty ambient water into a melt plume leads to the detrainment of the plume into the ambient (salt-)stratified ocean once it reaches neutral buoyancy \citep{magorrian_turbulent_2016,jackson_meltwater_2020}.
The flow we consider is turbulent due to the strong buoyancy forcing, and we are far from the marginal stability curves identified for this problem \citep{xin_bifurcation_1998,beaume_near-onset_2022}.

The rest of the paper is organised as follows.
We describe the governing equations, control parameters, and numerical methods used in section \ref{sec:setup}.
This is followed by presentation of the results where we highlight the effect of thermal buoyancy on the flow (\ref{sec:passive}), the global heat flux and how it is related to the salt flux (\ref{sec:global_flux}), the widths of the scalar boundary layers (\ref{sec:boundary_layers}), and the turbulent diffusivity away from the walls (\ref{sec:bulk_diff}).
Finally, we conclude and discuss our results in the context of ice-ocean interfaces in section \ref{sec:discussion}.

\section{Numerical methods and simulation setup \label{sec:setup}}

\begin{figure}
    \centering
    \includegraphics[width=\linewidth]{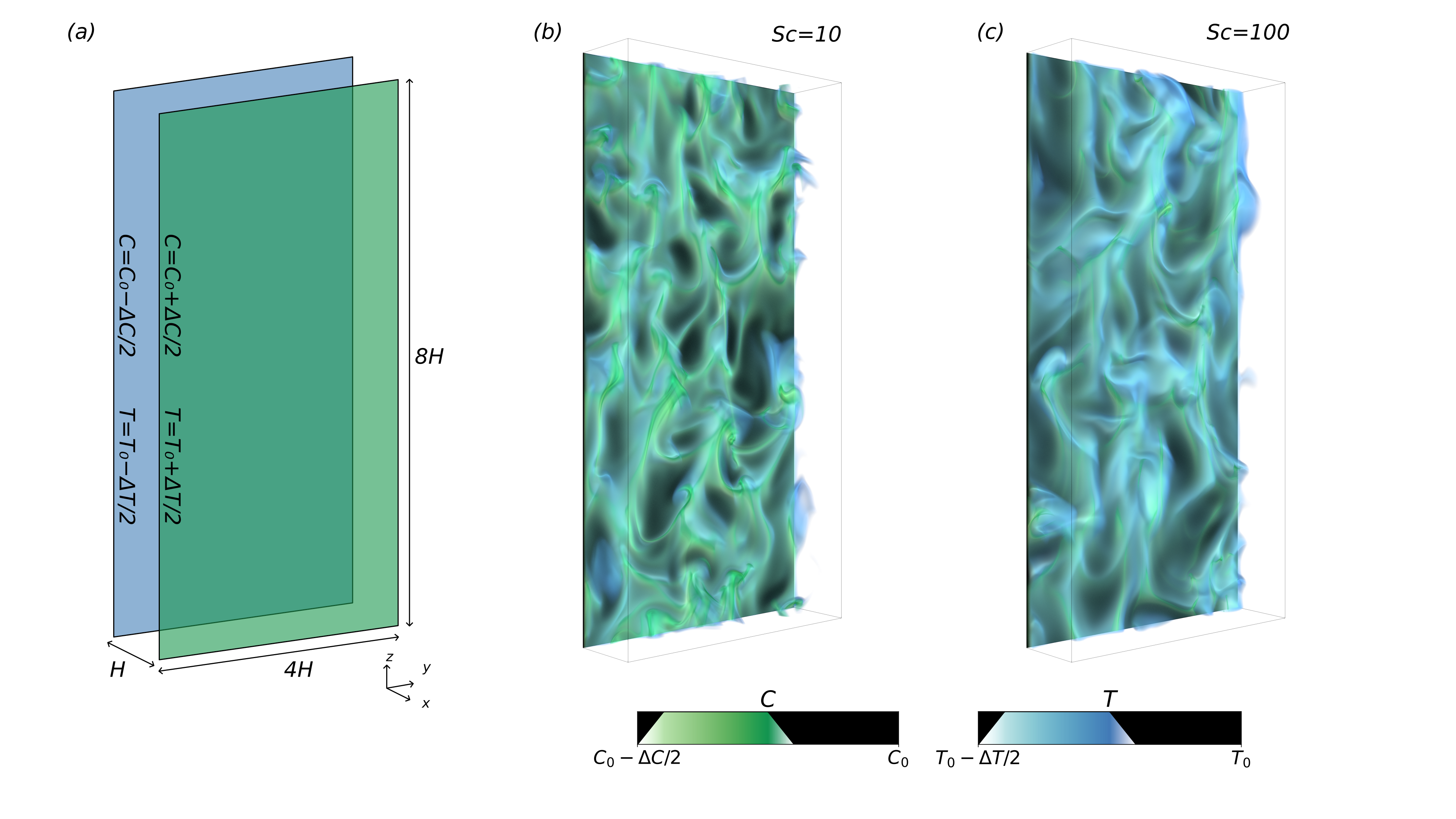}
    \caption{
        $(a)$ A schematic of the simulation domain, featuring two no-slip vertical planes; $(b,c)$ Volume renderings of the instantaneous temperature and salinity fields adjacent to the wall at $x=0$ from simulations $(b)$ A10L10 and $(c)$ A100L100.
        Colorbars show the opacity used for the volume rendering as well as the color, and the bounding box outlines the full extent of the domain.
        Slender, green plume structures highlight the buoyant regions of low salinity driving the flow up the wall, surrounded by more diffuse blue regions highlighting the low temperature patches.
        The buoyant flow due to the salinity perturbations advect these cold patches, so the structures in the two scalar fields become strongly correlated.
        With a higher $Sc$ and $Le$, the green salinity structures in $(c)$ are thinner and are nested more deeply in the diffuse temperature structures compared to those observed in $(b)$.
    }
    \label{fig:schematic}
\end{figure}

We consider the fluid flow inside a vertical channel of width $H$, with fixed values of temperature $T$ and solute concentration $C$ at each wall.
These impose a temperature difference $\Delta T$ and a concentration difference $\Delta C$ between the walls, where the density ratio $R_\rho=\beta_T \Delta T/\beta_C \Delta C = 0.02$ is fixed in all the simulations.
The oceanographic relevance of this value will be discussed later in \S \ref{sec:passive}.
Here, $\beta_T$ is the isobaric thermal expansion coefficient and $\beta_C$ is the haline contraction coefficient.
Following the Oberbeck--Boussinesq approximation, density differences obey a linear equation of state
\begin{equation}
    \rho = \rho_0\left( 1 - \beta_T (T - T_0) + \beta_C (C - C_0) \right) , \label{eq:EoS}
\end{equation}
and are only non-negligible in the buoyancy term of the momentum equations.
Real seawater has a nonlinear equation of state \citep{mcdougall_getting_2011}, and we later quantify errors associated with using the linear equation of state in \S \ref{sec:passive}.
We consider incompressible flow such that the velocity field $\bu$ satisfies $\grad \cdot \bu = 0$.
The temperature and concentration fields satisfy advection-diffusion equations, such that the full set of governing equations reads
\begin{align}
    \frac{\pd \bu}{\pd t} + (\bu \cdot \grad) \bu &= -{\rho_0}^{-1}\grad p + g(\beta_T T - \beta_C C) \mathbf{\hat{z}} + \nu \nabla^2 \bu , \label{eq:NSmom} \\
    \frac{\pd T}{\pd t} + (\bu \cdot \grad) T &= \kappa_T \nabla^2 T , \label{eq:Tevo}\\
    \frac{\pd C}{\pd t} + (\bu \cdot \grad) C &= \kappa_C \nabla^2 C . \label{eq:Cevo}
\end{align}
Here $g$ is gravitational acceleration, which acts in the $z$-direction, $\nu$ is the kinematic viscosity, and $\kappa_T$ and $\kappa_C$ are the molecular diffusivities of heat and salt respectively.
Values of these fluid properties relevant to the ocean are provided later in table \ref{tab:dimensional_quantities}.
We consider a domain of length $8H$ in the vertical direction ($z$) and $4H$ in the spanwise direction ($y$), and impose periodic boundary conditions along these axes, as in our previous single component study \citep{howland_boundary_2022}.
No slip boundary conditions ($\bu=0$) are applied at each wall, along with Dirichlet boundary conditions for the scalar field:
\begin{align}
    C &= C_0 - \Delta C/2, & T &= T_0 - \Delta T/2 & \textrm{at }x&=0 \\
    C &= C_0 + \Delta C/2, & T &= T_0 + \Delta T/2 & \textrm{at }x&=H.
\end{align}
A basic schematic of the domain is provided in figure \ref{fig:schematic}a.

Since there is no flow imposed in this system, its dynamics are uniquely determined by four dimensionless control parameters.
These are the aforementioned density ratio 
\begin{equation}
    R_\rho = \frac{\beta_T \Delta T}{\beta_C \Delta C},
\end{equation}
along with the Rayleigh number, Schmidt number, and Lewis number
\begin{align}
    Ra &= \frac{g\beta_C H^3 \Delta C}{\nu \kappa_C}, &
    Sc &= \frac{\nu}{\kappa_C}, &
    Le &= \frac{\kappa_T}{\kappa_C} .
\end{align}
We take the Rayleigh number to be based on the buoyancy of the concentration field since the density ratio is small and thus the buoyancy is mainly due to concentration differences.
Instead of the Rayleigh number, one could also characterise the dynamics in terms of the Grashof number $Gr=Ra/Sc$ which is equivalent to the square of a Reynolds number based on the free-fall velocity scale $U_f=\sqrt{g\beta_C H \Delta C}$ and the plate separation $H$.
Prescribing the Schmidt number and Lewis number in turn fixes the Prandtl number $Pr=\nu/\kappa_T$.

We solve the governing equations \eqref{eq:NSmom}-\eqref{eq:Cevo} numerically using our in-house Advanced Finite-Difference (AFiD) code.
Spatial derivatives are approximated by central second-order accurate finite differences, a Crank--Nicolson scheme is used to time-step the wall-normal diffusive terms, and a third-order Runge--Kutta scheme is used for all other terms following \citep{verzicco_finite-difference_1996,van_der_poel_pencil_2015}.
The slower diffusing scalar field is evolved on a higher resolution grid than the grid on which all other flow variables are stored.
We use tricubic Hermite interpolation between the two grids to compute the scalar advection and buoyancy terms following ref.\citep{ostilla-monico_multiple-resolution_2015}.
Grid stretching is used in the wall-normal direction to resolve the thin diffusive boundary layers, whereas grid spacing is uniform in the $y$ and $z$ directions.
Since the flow is anisotropic and dominated by thin plumes ejected from the boundary layers, the Batchelor scale is not a reliable estimate of the required resolution for each state variable \cite{shishkina_boundary_2010}.
We ensure resolution of the flow fields through a statistical convergence test, and by inspection of the power spectrum tails for both the velocity and scalar fields.

\begin{table}
    \caption{
        \label{tab:parameters}Input parameters and grid resolutions $N_x \times N_y \times N_z$ with domain size $H\times 4H \times 8H$ for the numerical simulations.
        Simulation names are constructed using the format A$X$L$Y$ for $Sc=X$, $Le=Y$ for the cases where temperature is an active (A) scalar, and the format P$X$ for $Sc=X$ for the cases where temperature is a passive (P) scalar.
    }
    \begin{ruledtabular}
        \begin{tabular}{lccccccc}
        Simulation  & $R_\rho$ & $Sc$ & $Ra$ & $Le$ & $Pr$ & Base grid resolution & Refined grid resolution \\
            A10L10  & 0.02  & 10    & $10^7$ & 10   & 1     & $192\times 512\times 1024$ & $384\times 1024\times 2048$\\
            A10L5   & 0.02  & 10    & $10^7$ & 5    & 2     & $192\times 512\times 1024$ & $384\times 1024\times 2048$\\
            A10L2   & 0.02  & 10    & $10^7$ & 2    & 5     & $192\times 512\times 1024$ & $384\times 1024\times 2048$\\
            A10L05  & 0.02  & 10    & $10^7$ & 0.5  & 20    & $256\times 768\times 1536$ & $384\times 1152\times 2304$\\
            A10L02  & 0.02  & 10    & $10^7$ & 0.2  & 50    & $256\times 768\times 1536$ & $512\times 1536\times 3072$\\
            A10L01  & 0.02  & 10    & $10^7$ & 0.1  & 100   & $256\times 768\times 1536$ & $512\times 2048\times 4096$\\
            A100L100& 0.02  & 100   & $10^8$ & 100  & 1     & $192\times 512\times 1024$ & $512\times 1536\times 3072$\\
            A100L50 & 0.02  & 100   & $10^8$ & 50   & 2     & $192\times 512\times 1024$ & $512\times 1536\times 3072$\\
            A100L20 & 0.02  & 100   & $10^8$ & 20   & 5     & $192\times 512\times 1024$ & $512\times 1536\times 3072$\\
            A100L10 & 0.02  & 100   & $10^8$ & 10   & 10    & $256\times 768\times 1536$ & $512\times 1536\times 3072$\\
            P10     & 0     & 10    & $10^7$ & 10   & 1     & $192\times 512\times 1024$ & $384\times 1024\times 2048$\\
            P100    & 0     & 100   & $10^8$ & 100  & 1     & $192\times 512\times 1024$ & $512\times 1536\times 3072$\\
        \end{tabular}
    \end{ruledtabular}
\end{table}

The input parameters for the numerical simulations are shown in table \ref{tab:parameters}.
We perform three sets of simulations, in which the Grashof number is always fixed at $Gr=10^6$.
This value is rather low compared to geophysical applications, but is sufficiently large to simulate turbulent convection.
At very large Grashof numbers, a transition from `buoyancy-driven' to `shear-driven' convection can be predicted \citep{wells_geophysical-scale_2008,malyarenko_synthesis_2020} where the fluxes follow a scaling associated with classical shear-driven turbulent boundary layers.
However, based on previous work \citep{wells_geophysical-scale_2008,ng_changes_2017,howland_boundary_2022}, it is expected that the boundary layers will not undergo this transition before other large scale phenomena such as ambient stratification or shear impact the dynamics.
In the first set of simulations, labelled A10 in table \ref{tab:parameters}, we fix $Sc=10$ and vary the Lewis number between 0.1 and 10.
We then fix $Sc=100$ for the second set (A100), varying the Lewis number between 10 and 100.
Finally, we consider two simulations (set P) where the density ratio is set to zero such that the temperature field is advected as a passive scalar.
Simulation A10L10 is initialised with linear temperature and salinity profiles with small-amplitude white noise to trigger the transition to turbulence before evolving to a statistically steady state.
All other simulations use the final state of that simulation as an initial condition to reduce the time needed to reach a steady state.
The simulations are each evolved for 300 free-fall time units ($H/U_f$) in this steady state, and any time-averaged results presented below are averaged over this period.

\section{Results \label{sec:results}}
\subsection{Thermal buoyancy effect \label{sec:passive}}

\begin{figure}
    \includegraphics{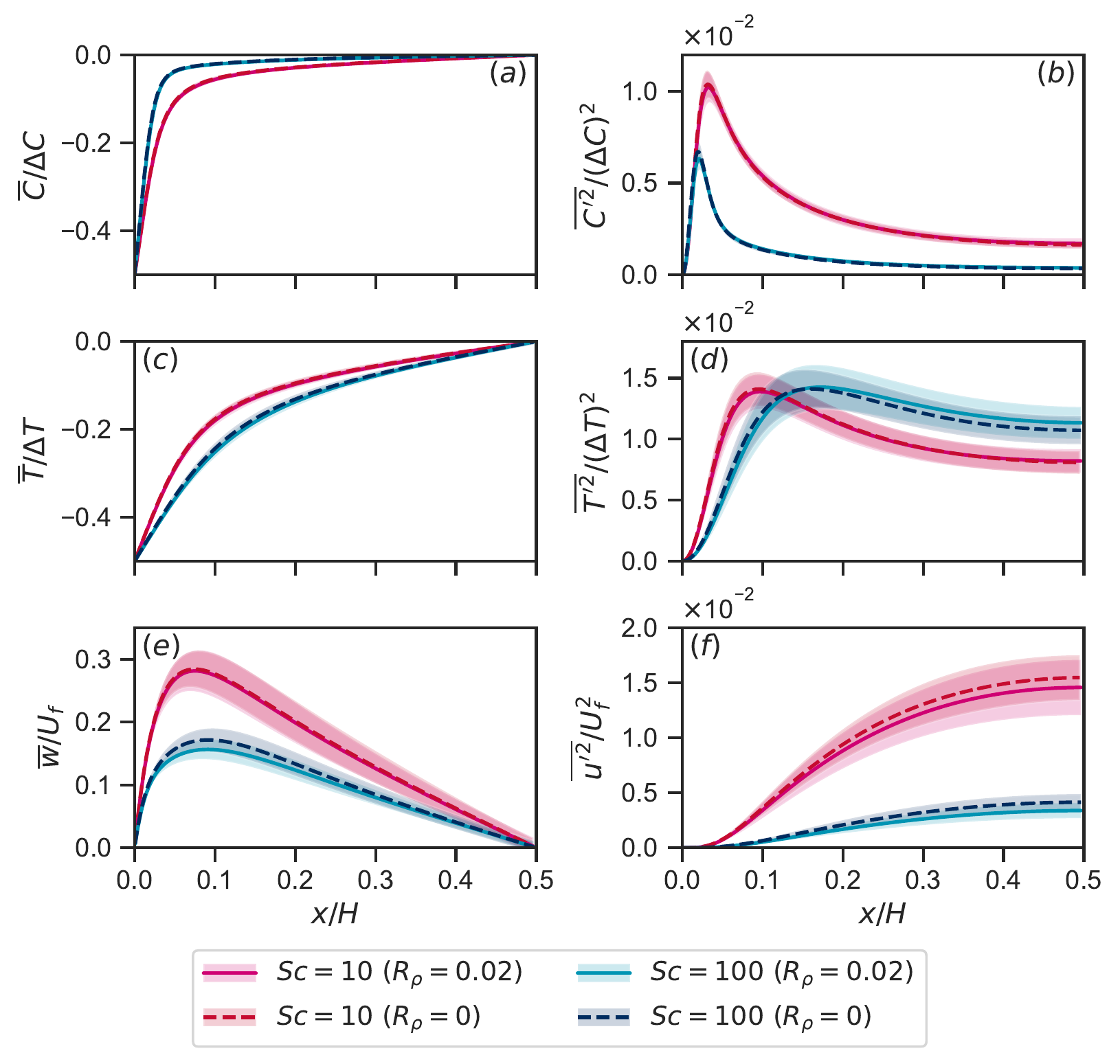}
    \caption{
        Wall-normal profiles of flow quantities from simulations P10 (red dashed), A10L10 (pink), P100 (blue dashed), and A100L100 (cyan).
        Solid lines show the time-averaged profiles and the shaded regions highlight the standard deviation (in time) of the $yz$-averaged profiles.
        Quantities plotted are $(a)$ mean concentration, $(b)$ concentration variance, $(c)$ mean temperature, $(d)$ temperature variance, $(e)$ mean vertical velocity, $(f)$ wall-normal velocity variance.
    }
    \label{fig:passive_comp}
\end{figure}

We begin our analysis by investigating whether the temperature field plays a significant role in the dynamics of the flow.
We compare the results of the set P simulations, where temperature is advected as a passive scalar ($R_\rho=0$), to their equivalent cases with $R_\rho=0.02$.
All of these cases have $Pr=1$ fixed.
In figure \ref{fig:passive_comp}, we compare various flow profiles between these simulations.
Here, we present profiles averaged in the vertical and spanwise directions, and then averaged in time, with the standard deviation in time of the mean profiles highlighted by shaded regions.
Visually, the cases with $R_\rho=0$ are very similar to those $R_\rho=0.02$ cases where temperature plays an active role in the buoyancy.

From the mean profiles, the most significant difference emerges in the vertical velocity for $Sc=100$, $Le=100$.
In this case the temperature field diffuses much more quickly than the concentration field, which is confined to a thin boundary layer.
Close to the vertical velocity peak, the contribution of salt to the buoyancy is reduced and so despite the small density ratio $R_\rho$, the temperature field can impact the vertical velocity through the buoyancy force.
A similar effect, although with a smaller impact, is observed in the $Sc=10$ case, where the vertical velocity peak is slightly reduced in the case of an active temperature field.
The reduction in peak velocity is also felt in the second order statistics plotted in the right column of figure \ref{fig:passive_comp}.
Both cases with an active temperature field exhibit a slight decrease in the wall-normal kinetic energy in the bulk when compared to the passive cases.
By contrast, the temperature variance in the bulk \emph{increases} when the thermal buoyancy component is included.
This may arise due to the marginally larger bulk temperature gradient in these simulations that would be in turn caused by the reduced mixing by the mean shear in the bulk.

Overall, although the effect of the thermal buoyancy component on the flow statistics is visible, it does not change the general picture describing the dynamics at $R_\rho=0.02$.
At this density ratio, the mean flow is driven by the buoyancy of the concentration field, and temperature is primarily transported as a passive scalar in this flow.
We note here that in terms of a realistic ice-ocean scenario, $R_\rho=0.02$ is even higher than what may be expected.
Ocean salinity has a typical concentration of \SI{35}{g.kg^{-1}}, with the value of concentration at the ice face set by the dynamic three-equation boundary condition \citep{martin_experimental_1977}.
\citet{kerr_dissolution_2015} performed experiments of a melting vertical ice face for ambient water temperatures between \SI{0.3}{\celsius} and \SI{5.4}{\celsius}, and estimated the interface salinity to vary between \SI{1.9}{g.kg^{-1}} and \SI{24.5}{g.kg^{-1}}.
Using their measurements and theoretical predictions, we can estimate $R_\rho$ by prescribing the haline contraction coefficient $\beta_C = \SI{7.86e-4}{(g.kg^{-1})^{-1}}$ and the thermal expansion coefficient $\beta_T = \SI{3.87e-5}{\kelvin^{-1}}$ from \citep{jenkins_convection-driven_2011}.
Although the true equation of state for seawater is nonlinear, and the effective thermal expansion coefficient varies with temperature, below \SI{5}{\celsius} these values are reasonable for seawater with a high concentration of salt.
Taking the temperature and salinity data from \citep{kerr_dissolution_2015}, we can compare the density computed with the linear equation of state \eqref{eq:EoS} against the fully nonlinear equation of state from the Gibbs SeaWater toolbox of TEOS-10 \citep{mcdougall_getting_2011}. We find relatively small errors of between $0.1\%$ and $1.5\%$ in the buoyancy forcing at the melting ice face, with the largest errors for the highest ambient temperatures.
Despite the varying far-field temperatures in the experiments of \citep{kerr_dissolution_2015}, the density ratio from their results remains roughly constant across all the experiments at $R_\rho\approx 8\times 10^{-3}$.
Given that our results show the heat transport is primarily passive at $R_\rho=0.02$, we expect this passive transport to also apply in oceanographically relevant flows.

\subsection{Flow visualization}

\begin{figure}
    \centering
    \includegraphics[width=\linewidth]{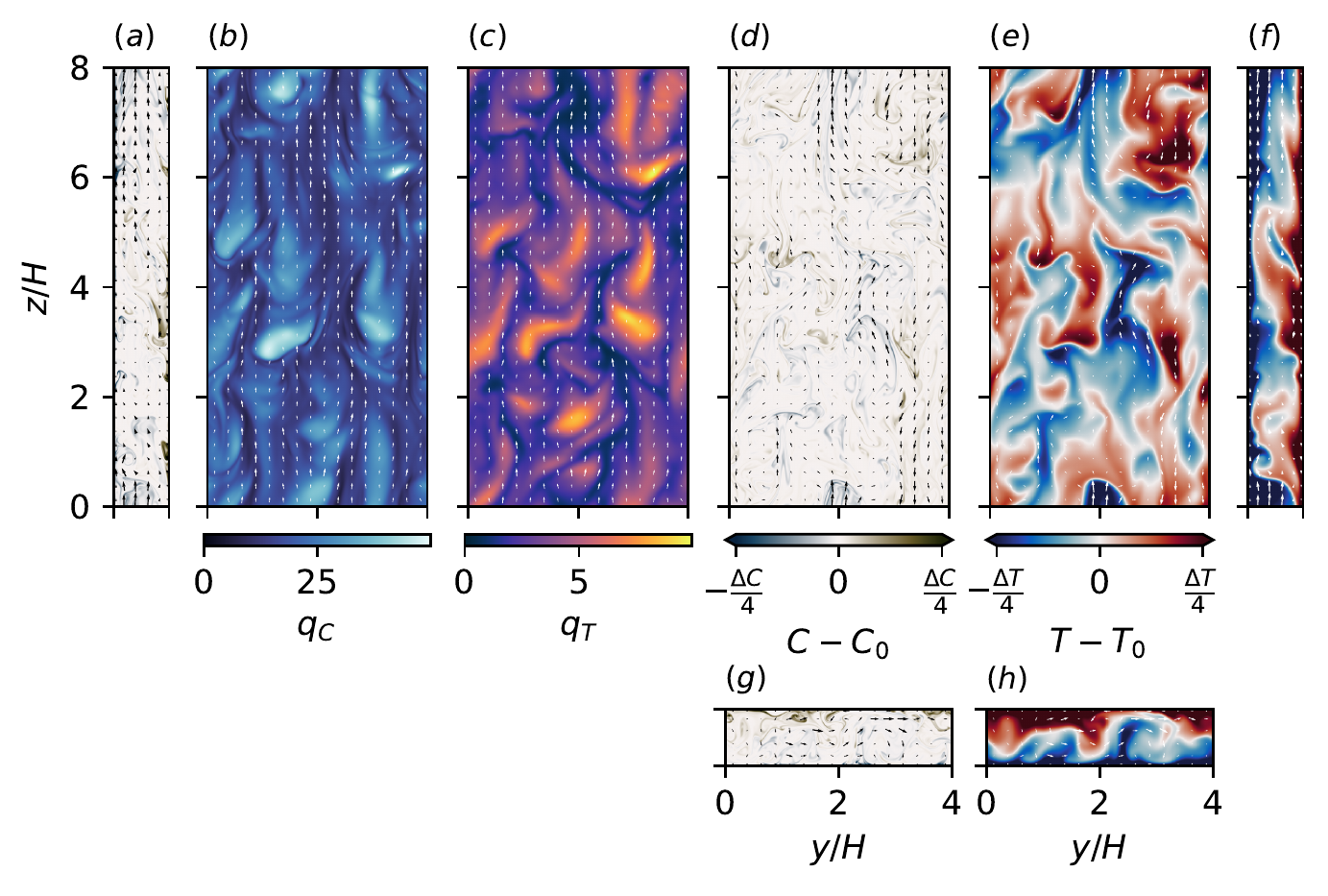}
    \caption{
        Instantaneous plane snapshots from simulation A100L100, where $Sc=100$, $Le=100$.
        In-plane velocity vectors are overlaid on each snapshot, which for the wall panels $(b,c)$ correspond to the local shear stress at the wall.
        $(a)$ Concentration field in the $xz$-plane $y=2H$;
        $(b,c)$ wall normal dimensionless flux of concentration and temperature at $x=0$ as defined in \eqref{eq:flux_def};
        $(d,e)$ concentration and temperature fields at the centre $yz$-plane $x=H/2$;
        $(f)$ temperature field in the $xz$-plane $y=2H$;
        $(g,h)$ concentration and temperature fields in the $xy$-plane $z=4H$.
        Since the concentration field drives the flow, regions of high or low concentration coincide with stronger vertical velocities.
        Structures in the temperature field are far more diffuse than those in the concentration field due to the large Lewis number.
    }
    \label{fig:planes}
\end{figure}

The passive role of the temperature field is highlighted visually by the volume renderings of figure \ref{fig:schematic}b-c. 
The green plumes representing the plumes of low salinity drive the buoyant flow, and carry with them perturbations of low temperature (shown by the light blue features).
The visual correlation is strong between the structures of the temperature field and the structures of the salinity field.
The faster diffusion of heat compared to salt is also visible in these renderings, with the blue temperature structures smoothed out relative to the thin, green plumes of low salinity.
This effect is amplified as the Lewis number increases, with the fresh perturbations in figure \ref{fig:schematic}c almost fully enveloped by the more diffuse temperature perturbations.

A more detailed snapshot of the flow dynamics is presented in figure \ref{fig:planes}, where we show both velocity and scalar fields in two-dimensional planes from the fully-developed statistically steady state of one simulation at $Le=100$, $Sc=100$.
The vertical planes of the salt concentration field (in panels a and d) highlight the correlation between vertical velocity and perturbations in salinity, due to the strong buoyancy driving provided by the concentration field.
As was the case for the 3-D visualizations, the temperature field in the plane snapshots (panels e, f, h) mimics the structures of the salinity field.
In the vertical planes (panels e and f), descending regions of warm fluid and rising regions of cool fluid highlight the negligible effect of temperature in driving the buoyant flow at $R_\rho=0.02$.
The usefulness of our multiple-resolution technique is also showcased by these panels, with the thin plumes of the concentration field requiring far finer resolution than the relatively diffuse temperature and velocity fields.

Panels b and c of figure \ref{fig:planes} focus on the near-wall dynamics, plotting the local dimensionless fluxes of salt and heat, defined as
\begin{equation}
    q_C = \frac{H}{\Delta C} \frac{\pd C}{\pd x}, \qquad
    q_T = \frac{H}{\Delta T} \frac{\pd T}{\pd x} . \label{eq:flux_def}
\end{equation}
The instantaneous local shear stress $\boldsymbol{\tau} = (\mu \partial_x v, \mu \partial_x w)$ is also plotted with arrows on the panels.
As observed in the single-component VC setup \citep{howland_boundary_2022}, the local shear stress is greatest in regions of \emph{low} local scalar flux.

\subsection{Global heat flux \label{sec:global_flux}}

Since heat is transported like a passive scalar in our simulations, the mean flow velocity and salt flux are solely determined by the Rayleigh and Schmidt numbers as in single-component vertical convection (VC).
The response of the VC system to these input parameters can be monitored in terms of the Nusselt number (considered here for salt), the Reynolds number, and the shear Reynolds number:
\begin{equation}
    Nu_C = \frac{H F_C}{\kappa_C \Delta C} = \frac{H}{\Delta C} \left.\frac{\pd \overline{C}}{\pd x}\right|_\textrm{wall} = \overline{q_C} , \qquad
    Re = \frac{\max \overline{w} H}{\nu} , \qquad
    Re_\tau = \frac{V_\ast H}{\nu} ,
\end{equation}
where $F_C$ is the diffusive salt flux at the wall and $V_\ast = \sqrt{\tau_\textrm{w}/\rho}$ is the friction velocity based on the measured wall shear stress ${\tau_\textrm{w} = \mu (\pd_x \overline{w})_\textrm{wall}}$.
The simulations described in this manuscript all closely follow our previous results for single-component VC \citep{howland_boundary_2022}, where we observed power-law relations of
\begin{equation}
    Nu_C\propto Ra^{0.321} Sc^{-0.083} , \qquad
    Re\propto Ra^{0.489} Sc^{-0.738} , \qquad
    Re_\tau \propto Ra^{0.362} Sc^{-0.446} ,
\end{equation}
from a two-parameter regression for the parameter range $10^6\leq Ra \leq 10^9$, $1 \leq Sc \leq 100$.
This best-fit power-law description is rather simplistic, and at yet higher $Ra$ one can reasonably expect a regime change to a fully shear-driven boundary layer \citep{ng_changes_2017}.

We characterise the heat flux in terms of the thermal Nusselt number, defined
\begin{equation}
    \label{eq:Nusselt}
    {Nu}_T = \frac{H F_T}{\kappa \Delta T} = \frac{H}{\Delta T} \left.\frac{\pd \overline{T}}{\pd x}\right|_{\textrm{wall}} ,
\end{equation}
where $F_T$ is the mean diffusive heat flux at the walls.
When investigating the effect of differential diffusion on the heat flux, it is useful to consider the heat flux in terms of its ratio to the salt flux.
If the Lewis number were equal to one, so heat and salt diffused at the same rates, then the governing equations \eqref{eq:Tevo} and \eqref{eq:Cevo} would become identical, and the heat and salt fluxes would therefore be equal.
We can thus investigate the dependence of the ratio
\begin{equation}
    R=\frac{Nu_C}{Nu_T} = \frac{\Delta T (\partial_x \overline{C})_\mathrm{wall}}{\Delta C (\partial_x \overline{T})_\mathrm{wall}} \label{eq:ratio_def}
\end{equation}
on the Lewis number.
This quantity is sometimes referred to as the temperature-to-salt boundary layer ratio \citep{rosevear_role_2021} since $H/(2Nu)$ is a commonly used measure of scalar boundary layers.
Some ice-ocean studies instead refer to the dimensionless flux ratio $\gamma=F_T \Delta C/F_C \Delta T$, which is directly related to $R$ through $\gamma R=Le$ \citep{notz_impact_2003}.
Both $\gamma$ and $R$ are equal to one when $Le=1$.

\begin{figure}
    \includegraphics{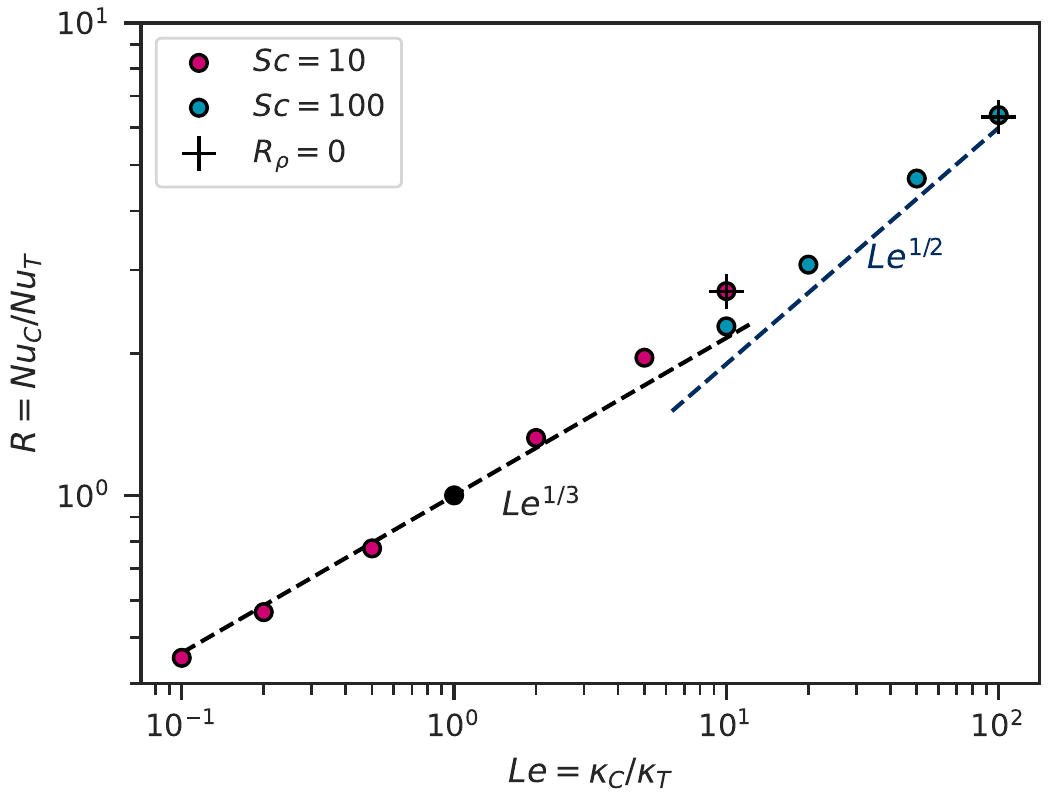}
    \caption{
        Ratio $R$ of solutal Nusselt number $Nu_C$ to thermal Nusselt number $Nu_T$ across all the simulations plotted against the Lewis number.
        Colours denote the Schmidt number of the simulations, and crosses are used to highlight the cases with zero density ratio.
        The dashed straight lines with slopes $1/3$ and $1/2$ are shown for comparison.
        The black dot highlights the fixed theoretical point $R=Le=1$.
    }
    \label{fig:flux_ratio}
\end{figure}

We plot the Nusselt number ratio $R$ measured from our simulations against the Lewis number in figure \ref{fig:flux_ratio}a.
The passive temperature cases of set P overlay the active cases near-perfectly.
For low Lewis numbers $Le \leq 1$, we find good agreement with a $1/3$ scaling law.
In the following paragraph, we will show how this can be explained in conjunction with our previous finding in \citep{howland_boundary_2022} that the scalar flux in vertical convection at moderate Rayleigh number and high Schmidt number is consistent with $Nu_C \sim Re_\tau Sc^{1/3}$.
Here, $Re_\tau = V_\ast H/\nu$ is the shear Reynolds number, calculated from the friction velocity $V_\ast=\sqrt{\tau_w/\rho}$ based on the measured wall shear stress $\tau_w=\mu (\partial_x \overline{w})_\mathrm{wall}$.
Such a scaling with $Sc$ is more widely applicable in high Schmidt number turbulent boundary layers \cite{kader_heat_1972-1}, where the $Sc^{1/3}$ factor arises due to the scalar boundary layer being nested within the viscous sublayer \citep{schlichting_boundary-layer_2016}.

Following section 9.3 of \citet{schlichting_boundary-layer_2016}, we can explicitly formulate the boundary layer equations for a passive scalar (written for temperature below) as
\begin{equation}
    w = \frac{\tau}{\mu} x, \qquad w\frac{\pd T}{\pd z} = \frac{\tau}{\mu} x \frac{\pd T}{\pd z} = \kappa \frac{\pd^2 T}{\pd x^2} . \label{eq:BL_eqn}
\end{equation}
Here, we assume that the dominant balance in the boundary layer is between wall-normal diffusion and advection by the mean shear, which is uniform within the viscous sublayer.
The boundary layer equation \eqref{eq:BL_eqn} permits a similarity solution under substitution of the similarity variable $\eta=(\tau/3\mu\kappa)^{1/3}x z^{-1/3}$.
Under this transformation, the Prandtl number dependence emerges as $Nu\sim Pr^{1/3}$.
Since temperature acts as a passive scalar in our simulations and $Pr\geq1$ in all the cases we consider, we may expect an equivalent relationship as $Nu_T \sim Re_\tau Pr^{1/3}$.
In this case, the Nusselt number ratio is $R\sim Le^{1/3}$.
Since this scaling argument is consistent with the single-component VC results of \cite{howland_boundary_2022}, it should be valid at $Le=1$, where we know that $R=1$.
For the scaling's range of validity, there should therefore be no pre-factor and we get $R=Le^{1/3}$.

However, as $Le$ increases, the data deviates from the $1/3$ slope and the trend becomes steeper.
This is most evident in the simulations with $Sc=100$, where the effective scaling exponent of the data begins to approach $1/2$.
Such a scaling has been used previously for convective boundary layers at vertical ice faces in \citep{kerr_dissolution_2015}.
An argument for this scaling was provided by \citet{kerr_dissolving_1994}, who considered the boundary layers at a horizontal ice face driving solutal convection in salt water above.
In this scenario, the thermal and solutal boundary layers grow diffusively until the solutal boundary becomes convectively unstable.
We neglect the influence of shear; then the two boundary layer widths satisfy
\begin{equation}
    \delta_T \sim \sqrt{\kappa_T t} , \qquad
    \delta_C \sim \sqrt{\kappa_C t} ,
\end{equation}
during the diffusive growth phase.
Taking the Nusselt numbers to be inversely proportional to the boundary layer widths, we find that
\begin{equation}
    R=\frac{Nu_T}{Nu_C} \sim \frac{\delta_C}{\delta_T} \sim \sqrt{\frac{\kappa_C}{\kappa_T}} = Le^{1/2} .
\end{equation}
Due to the buoyancy driving of the concentration field, the solutal boundary layer becomes convectively unstable as it reaches a certain width.
This instability whips both boundary layers away from the wall, after which the diffusive growth of new boundary layers restarts at the wall.
With the total fluxes governed by this process of diffusive growth intermittently reset by convective instabilities, 
\citet{kerr_dissolving_1994} concludes that the Nusselt number ratio must therefore scale as $Le^{1/2}$.

The key difference in assumptions between the two observed scalings is whether the thermal boundary layer is nested within the viscous sublayer.
If this is the case, the entire thermal boundary layer experiences a velocity field of approximately uniform shear and the subsequent similarity solution gives a $R=Le^{1/3}$ result.
If the thermal boundary layer is not nested, then it can diffuse essentially unaffected by the shear, such that a diffusive $R\sim Le^{1/2}$ result holds.
Since the diffusive boundary layers are so important to these scaling arguments, we directly inspect them in the following section to gain more insight on the transition between these scaling regimes.

\subsection{Boundary layer analysis \label{sec:boundary_layers}}

To investigate whether the above arguments are suitable for describing the heat flux through the system, we now explicitly analyse the scalar boundary layers in the simulations.
We define the width of the thermal boundary layer as follows in terms of the nature of the heat flux.
Taking a $yz$-average of the advection-diffusion equation \eqref{eq:Tevo}, assuming a statistically steady state, and integrating with respect to $x$ shows us that the mean heat flux is uniform across all wall-normal locations:
\begin{equation}
    F_T(x) = \underbrace{\kappa_T \frac{\pd \overline{T}}{\pd x}}_\textrm{diffusive heat flux} + \underbrace{\overline{-u'T'}}_\textrm{turbulent heat flux} = \mathrm{constant} .
    \label{eq:heat_flux}
\end{equation}
Here an overbar denotes an average with respect to $y$, $z$, and $t$, and a prime denotes the perturbation from this average.
Far from the walls, the heat flux is dominated by its turbulent component and the mean temperature gradient is small.
However, due to the no-penetration condition at the walls, heat flux at the boundaries must be purely diffusive.
We therefore define the thermal boundary layer as the region where the diffusive flux is the dominant contribution to the heat flux.
The boundary layer width is then defined as the crossover location of the fluxes:
\begin{equation}
    \delta_T = x|_{\kappa \partial_x \overline{T} = -\overline{u'T'}}.
\end{equation}

\begin{figure}
    \includegraphics{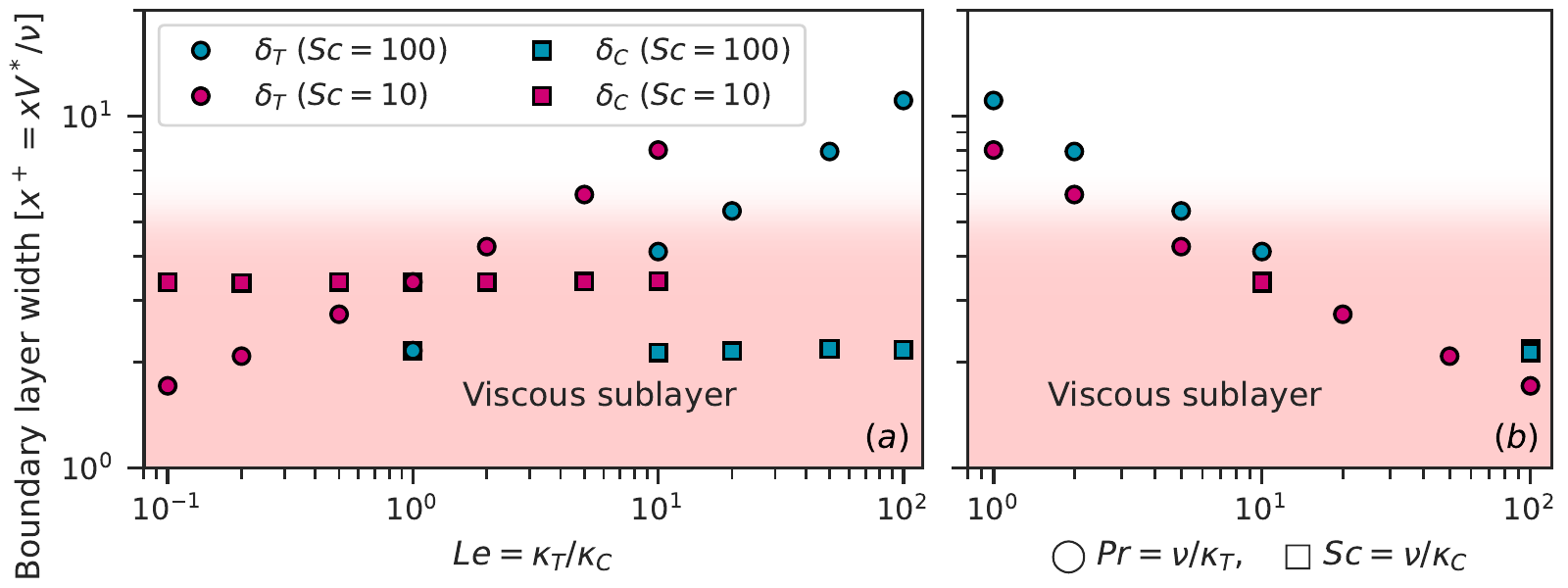}
    \caption{
        Thermal (circles) and solutal (squares) boundary layer widths in viscous units plotted against $(a)$ the Lewis number and $(b)$ the Prandtl or Schmidt number.
        The viscous sublayer, defined as $x^+\lesssim 5$, is highlighted by the red shading.
        Data points for $Le=1$ in $(a)$ are inferred from the solutal boundary layer widths for each dataset (A10 and A100).
        For $Le=1$ the thermal and solutal boundary layer widths must be equal.
    }
    \label{fig:BL_widths}
\end{figure}

In figure \ref{fig:BL_widths} we plot the thermal boundary layer width in terms of the viscous wall unit $x^+ = x V^*/\nu$, where $V^*$ is the friction velocity calculated from the mean shear stress at the wall.
For the region $x^+=O(1)$, viscous forces are dominant, and it is common to define the viscous sublayer as $x^+\lesssim 5$ in turbulent flows \citep{davidson_turbulence:_2015}.
This sublayer is highlighted by the red shaded region in figure \ref{fig:BL_widths}.
Here, we also plot the solutal boundary layer width $\delta_C$, which is defined in an analogous way to the thermal boundary layer width.
The solutal boundary layer width $\delta_C$ is unaffected by the Lewis number, providing further evidence for the passive role of the temperature field.
In both sets of simulations, $\delta_C^+<5$, i.e. the solutal boundary layer is nested within the viscous sublayer.
As mentioned in the previous subsection, a nested scalar boundary layer is consistent with the scaling $Nu\sim Re_\tau Sc^{1/3}$, and so figure \ref{fig:BL_widths} provides some insight into why the data of \citep{howland_boundary_2022} agrees with that relationship.
For low $Le$, the thermal boundary layer is thinner than the solutal boundary layer and is therefore also nested within the viscous sublayer.
When both thermal and solutal boundary layers are nested within the viscous one, we anticipate the flux ratio $R = Le^{1/3}$ observed for low $Le$ in figure \ref{fig:flux_ratio}a.

Figure \ref{fig:BL_widths}a also provides insight on why the deviations from the $Le^{1/3}$ flux ratio in figure \ref{fig:flux_ratio}a occur at different values of $Le$ for the two different Schmidt numbers.
For the Grashof number $Gr=10^6$ considered in this study, the solutal boundary layer for $Sc=100$ is nested deeper within the viscous sublayer than for $Sc=10$.
Assuming that $R=Le^{1/3}$ applies whenever both scalar boundary layers satisfy $\delta^+<5$, the Lewis number at which the thermal boundary layer reaches the edge of the viscous sublayer must therefore be larger for $Sc=100$.
Only once $\delta_T^+>5$ will we see a deviation from the $Nu_T\sim Re_\tau Pr^{1/3}$ scaling.

In figure \ref{fig:BL_widths}b, we plot the same sublayer thickness data against the Prandtl number (or Schmidt number for solutal boundary layers).
Here, we see that both scalar boundary layers primarily depend on $Pr$ or $Sc$, rather than $Le$.
From this, we can provide a rough estimation of $Pr\approx 10$ as the critical value above which the thermal boundary layer is nested within the viscous sublayer.
We discuss the range of applicability of this criterion later in \S \ref{sec:discussion}.


\subsection{Turbulent diffusivity in the bulk \label{sec:bulk_diff}}

\begin{figure}
    \includegraphics{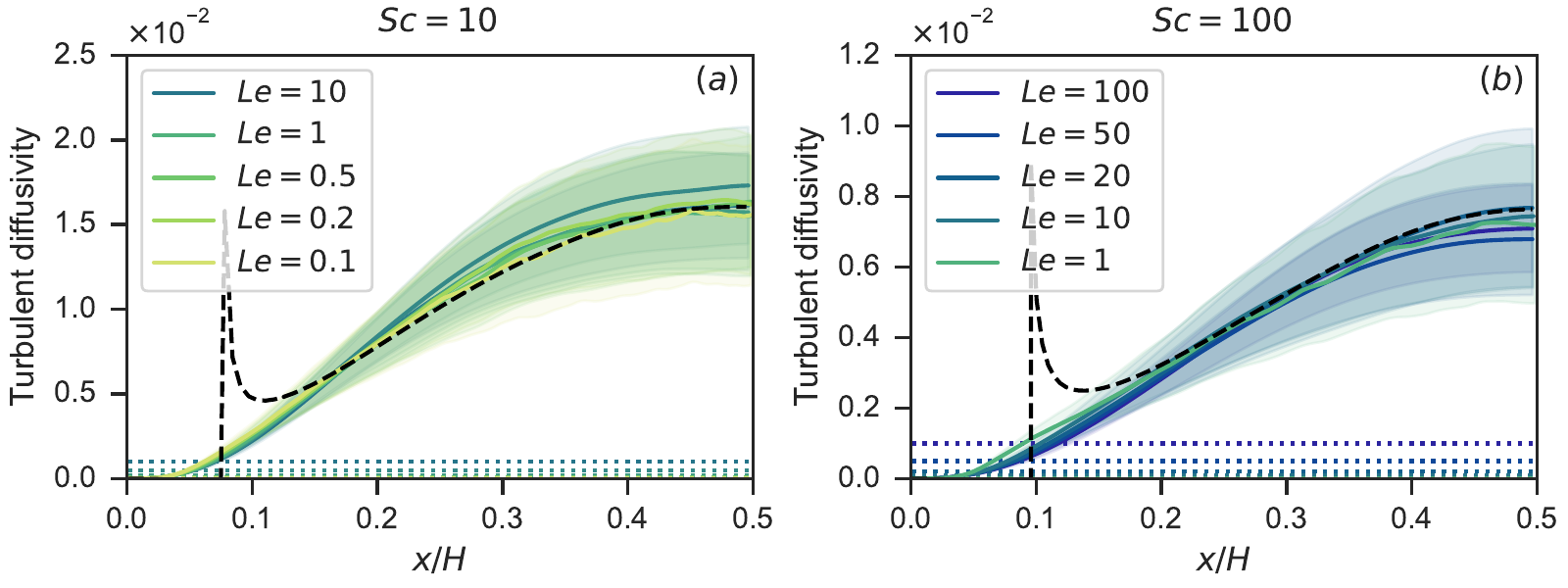}
    \caption{
        Wall-normal profiles of the turbulent heat diffusivity $K_T$ for each simulation, for two different Schmidt numbers, $(a)$ $Sc=10$ and $(b)$ $Sc=100$.
        Curves labelled as $Le=1$ represent the turbulent diffusivity of the concentration field.
        For comparison, the wall-normal profile of the turbulent viscosity $\nu_t$ is also plotted as a dark dashed line, and the constant values of molecular diffusivity $\kappa_T$ are plotted as dotted lines with colours matching the legend.
        Shaded regions highlight the temporal standard deviation of the turbulent heat flux $\overline{u'T'}$, normalised by the mean temperature gradient $\pd_x\overline{T}$ (multiplied by $\kappa_T$).
    }
    \label{fig:turb_diff}
\end{figure}

We conclude our analysis of the heat transport in the simulations by investigating the behaviour of the turbulent bulk away from the walls.
In this region, the heat flux defined in \eqref{eq:heat_flux} is dominated by the turbulent contribution.
To gain insight into the transport properties of the bulk, we can rewrite \eqref{eq:heat_flux} in terms of a turbulent diffusivity $K_T(x)$, such that
\begin{align}
    K_T(x) &= -\frac{\overline{u'T'}}{\partial_x \overline{T}}, &
    F_T(x) &= \left( \kappa_T + K_T(x) \right) \frac{\pd \overline{T}}{\pd x} .
\end{align}
We plot the wall-normal profiles of $K_T(x)$ for each of the simulations in sets A10 and A100 in figures \ref{fig:turb_diff}a and \ref{fig:turb_diff}b respectively.
For each set, an equivalent profile of the turbulent salt diffusivity $K_C(x)=-\overline{u'C'}/\partial_x \overline{C}$ is also plotted for comparison and labelled as $Le=1$.
In all the simulations, we observe that the turbulent diffusivities are far greater than the molecular diffusivities away from the walls, and that the Lewis number has no significant effect on the profile of $K_T(x)$ in the bulk.
The turbulence in the bulk thus mixes the temperature and concentration fields at an equal rate, and there are no double-diffusive effects on the mean profiles.

Furthermore, we find that the wall-normal momentum transport is also approximately equal to the scalar transport in the bulk.
We quantify this by calculating the turbulent viscosity $\nu_t(x)=-\overline{u'w'}/\partial_x \overline{w}$, and also plotting it in figure \ref{fig:turb_diff} as black, dashed lines.
Recall that temperature acts as a passive scalar, so the turbulent viscosity and turbulent salt diffusivity profiles will be identical for all the simulations with the same $Sc$.
We therefore only plot one profile on each panel for these quantities.
The turbulent viscosity is negative close to the walls, and becomes ill-defined at the velocity extrema, so using a simple model based on a turbulent viscosity would be inappropriate for describing the mean evolution of this vertical convection flow.
Nevertheless, $\nu_t$ agrees rather nicely with $K_T$ in the bulk away from the velocity maximum.
The heat and salt transport in the bulk therefore satisfy $Pr_t\approx 1$ and $Sc_t\approx 1$, where $Pr_t=\nu_t/K_T$ and $Sc_t=\nu_t/K_C$ are the turbulent Prandtl number and turbulent Schmidt number.
Indeed, convergence to $Pr_t\approx 1$ in uniformly sheared flow regions away from boundaries is frequently observed in a range of flows coupling shear and buoyancy effects \citep{chung_direct_2012,portwood_asymptotic_2019,van_reeuwijk_mixing_2019}.

\section{Discussion and conclusions \label{sec:discussion}}

In this study, we have investigated the effect of differential diffusion on the transport of heat and salt through a multicomponent fluid in vertical convection.
For a density ratio of $R_\rho=0.02$, relevant to the meltwater-driven convection at a vertical ice face in the ocean, we find that the convection is driven by differences in salt concentration, and that the contributions of the temperature to the buoyancy forcing are insignificant.
Through comparison with the case of $R_\rho=0$, where temperature is advected as a passive scalar, we conclude that classical double-diffusive phenomena such as salt fingers or diffusive convection are largely irrelevant in this flow geometry.
This is further evidenced by the independence of the turbulent heat diffusivity in the bulk on the Lewis number.
The heat transport away from the walls is characterised by a turbulent Prandtl number of $Pr_t\approx 1$, meaning that heat, salt, and momentum are all mixed at the same rate.
We therefore do not expect double-diffusive convection to play a significant role in the flow dynamics at vertical ice faces.

However, the difference in the molecular diffusivities of heat and salt, characterised by the Lewis number $Le=\kappa_C/\kappa_T$, is important in determining the relative fluxes of heat and salt through the system.
When $Le<1$, the ratio of salt flux to heat flux satisfies $R = Le^{1/3}$, but a steeper trend emerges as the Lewis number increases towards realistic values for salt water.
This increase can be explained by the relative widths of the diffusive and viscous sublayers and their dependence on the Lewis number.
Whenever both scalar boundary layers, defined as the regions where diffusive flux is larger than turbulent flux, are nested within the viscous sublayer, the scalar fluxes follow the classical high $Pr$ scaling $Nu\sim Re_\tau Pr^{1/3}$ and the flux ratio therefore satisfies $R=Le^{1/3}$.
As $Le$ increases, the thermal sublayer can extend beyond the edge of the viscous sublayer, causing this prediction to break down.
In this case the effective scaling exponent grows towards $1/2$, as suggested by \citep{kerr_dissolving_1994} for diffusing boundary layers intermittently shed by instabilities.


\begin{figure}
    \includegraphics{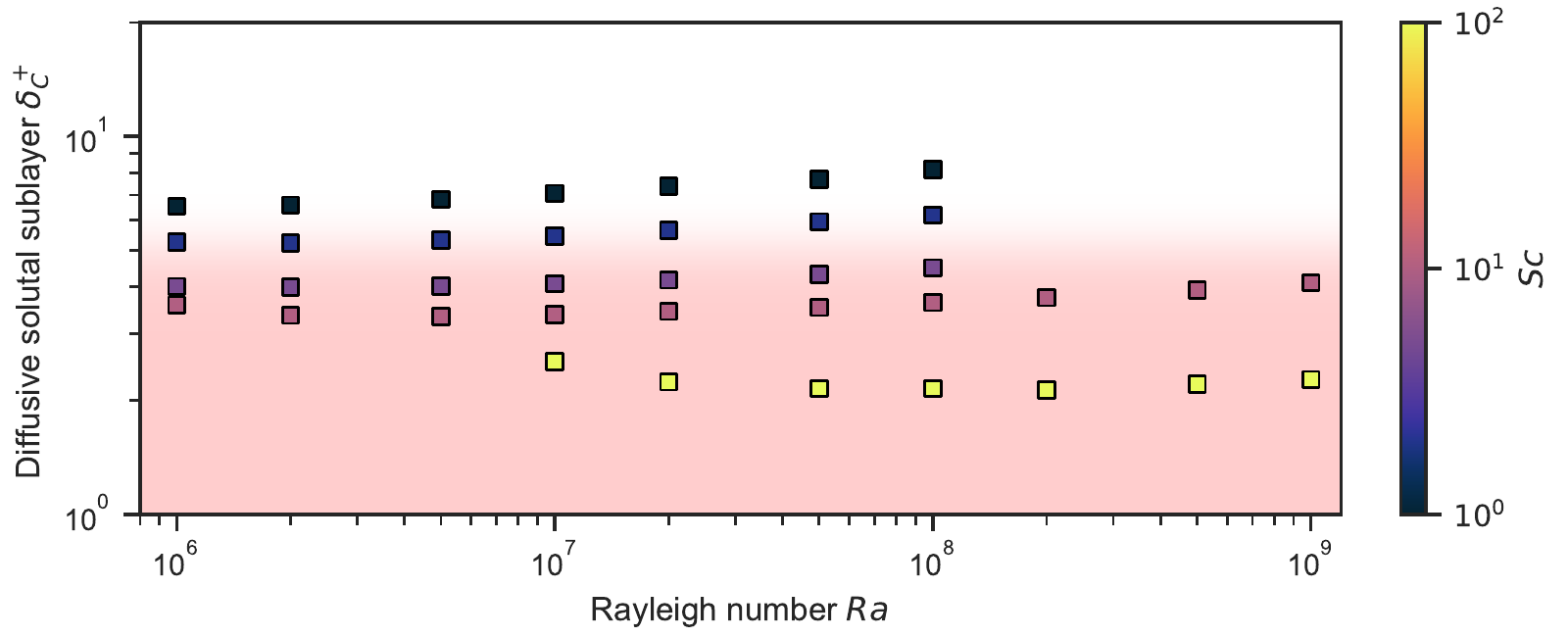}
    \caption{
        Data from \citep{howland_boundary_2022} showing the width of the diffusive solutal sublayer in viscous units ${\delta_C^+=\delta_C V^*/\nu}$. The viscous sublayer $x^+\lesssim 5$ is shaded in the same manner as in figure \ref{fig:BL_widths}.
        The solutal sublayer width depends primarily on $Sc$, with a very slight increase observed with $Ra$.
    }
    \label{fig:solutal_sublayer}
\end{figure}

Figure \ref{fig:BL_widths}b highlights that the transition between scaling relations can be roughly estimated by a critical Prandtl number of $Pr_c\approx 10$.
All the simulations performed in this study have had fixed Grashof number $Gr=10^6$, but we can infer the wider applicability of this result by consulting data from our previous work on high $Sc$ vertical convection \citep{howland_boundary_2022}.
In figure \ref{fig:solutal_sublayer}, we plot the width of the diffusive solutal boundary layer $\delta_C^+$ in viscous wall units for a wider range of $10^6\leq Ra\leq 10^9$ and $1\leq Sc\leq 100$.
Although there is some small variation with $Ra$, the dominant variation is associated with changing $Sc$, following the same trend as seen in figure \ref{fig:BL_widths}b for $\delta_T^+$ as a function of $Pr$.
Assuming that $Pr_c\approx 10$ is an appropriate approximation for the $R$-$Le$ scaling transition for varying $Sc$ and $Ra$, we arrive at a useful result for real ice-ocean systems where $Pr=14$.
Given these results, the flux ratio $R=Le^{1/3}$ appears most suitable for application to vertical ice faces.
One caveat to this is the slight increase in $\delta_C^+$ with $Ra$ observed with higher $Ra$.
This suggests that the critical $Sc$ (and hence $Pr$) at which the scalar boundary layer reaches the edge of the viscous sublayer will gradually increase as $Ra$ increases.
Nevertheless, as long as $Pr_c$ remains close to 14, $R=Le^{1/3}$ will be the most appropriate prediction for the flux ratio.

More generally, for the parameter range considered in this study, $Le^{1/3}$ appears to be a physical lower bound for the Nusselt number ratio $R$ in low-density ratio vertical convection with heat and salt.
Extrapolating the two dashed lines from figure \ref{fig:flux_ratio} out to a typical Lewis number for polar oceans of $Le=204$ \citep{notz_impact_2003} gives a range of $5.88<R<8.57$ in the current study.
The lower of these values, consistent with the shear-driven $Le^{1/3}$ prediction, is equivalent to a flux ratio of $\gamma=F_T/F_C= 35$ used in common ice-ocean parameterisations \citep{holland_modeling_1999,jenkins_convection-driven_2011}.
This contrasts somewhat to previous results for diffusive convection underneath \emph{horizontal} ice surfaces, where lower values of $R$ are often inferred or theorized.
For the same Lewis number, \citet{notz_impact_2003} develop a theory describing the ablation of `false bottoms' on the underside of ice floes, where the Nusselt number ratio is predicted to lie in the range $2.92<R<5.84$.
Numerical studies of diffusive convection for varying Lewis numbers have shown that the dependence of $R$ on $Le$ appears to decrease as $Le$ increases \citep{keitzl_reconciling_2016}, contrary to our findings in figure \ref{fig:flux_ratio}.
In diffusive convection, the solutal boundary layer plays a stabilising role, since the fresh layer overlies the saltier ambient.
Motion is suppressed in this sublayer, meaning heat flux is purely diffusive there, and the outer flow is driven purely by instability of the thermal boundary layer outside the diffusive solutal boundary layer.
The instability restricts the growth of the thermal boundary layer, so it does not become much thicker than the diffusive solutal boundary layer.
By contrast, in vertical convection it is the thinner solutal boundary layer that drives the motion, and the thermal boundary layer acts passively.

Finally, we can consider how significant these discrepancies in the flux ratio can be for predictions of the melt rate at an ice-ocean interface.
We use the three-equation boundary condition due to the salt-dependence of the melting point, and the conservation of heat and salt (repeated from \eqref{eq:ice_BC}):
\begin{align}
    T_i + \lambda C_i &= 0, &
    \frac{L}{c_p} \mathcal{V} &= F_T, &
    C_i \mathcal{V} &= F_C . \label{eq:melt_BC}
\end{align}
As a reminder, $T_i$ and $C_i$ are the values of temperature and salt concentration at the ice-water interface, $\lambda$ is the liquidus slope, $L$ is the latent heat, $c_p$ is the specific heat capacity, and $\mathcal{V}$ is the ablation velocity, or melt rate, of the ice.
For simplicity, we consider the ice to be isothermal such that conduction in the solid is zero.
Although latent heat is not considered directly in our simulations, from the second condition of \eqref{eq:melt_BC} we note that the latent heat simply contributes to a constant scaling factor between the melt rate to the heat flux.
If $L/c_p$ is large, the melt rate is slow relative to the dynamics of the flow and we can assume our results for a stationary boundary will be relevant to the case of an evolving planar boundary.
From \eqref{eq:melt_BC}, we can deduce that the Nusselt number ratio $R\propto F_C/F_T$, as defined in \eqref{eq:ratio_def}, determines the interfacial salinity $C_i$ through the quadratic equation
\begin{equation}
    \frac{L}{c_p} (C_\infty - C_i) = \frac{Le}{R} C_i (T_\infty + \lambda C_i) ,
\end{equation}
where $T_\infty$ and $C_\infty$ are the far-field values of temperature and concentration.

\begin{table}
    \caption{\label{tab:dimensional_quantities} Physical quantities related to the ice-water boundary condition \citep{jenkins_convection-driven_2011}.}
    \begin{ruledtabular}
        \begin{tabular}{ccccccc}
            $\lambda$ [\si{\kelvin (g.kg^{-1})}] & $L/c_p$ [\si{\kelvin}] & $C_\infty$ [\si{g.kg^{-1}}] & $\nu$ [\si{m^2s^{-1}}] & $\kappa_T$ [\si{m^2 s^{-1}}] & $\kappa_C$ [\si{m^2 s^{-1}}] & $\beta_C$ [\si{(g.kg^{-1})^{-1}}]\\
            $5.73\times10^{-2}$ & $84.0$ & $34.5$ & $1.95\times 10^{-6}$ & $1.41\times 10^{-7}$ & $8.02\times 10^{-10}$ & $7.86\times 10^{-4}$\\
        \end{tabular}
    \end{ruledtabular}
\end{table}
\begin{figure}
    \centering
    \includegraphics[]{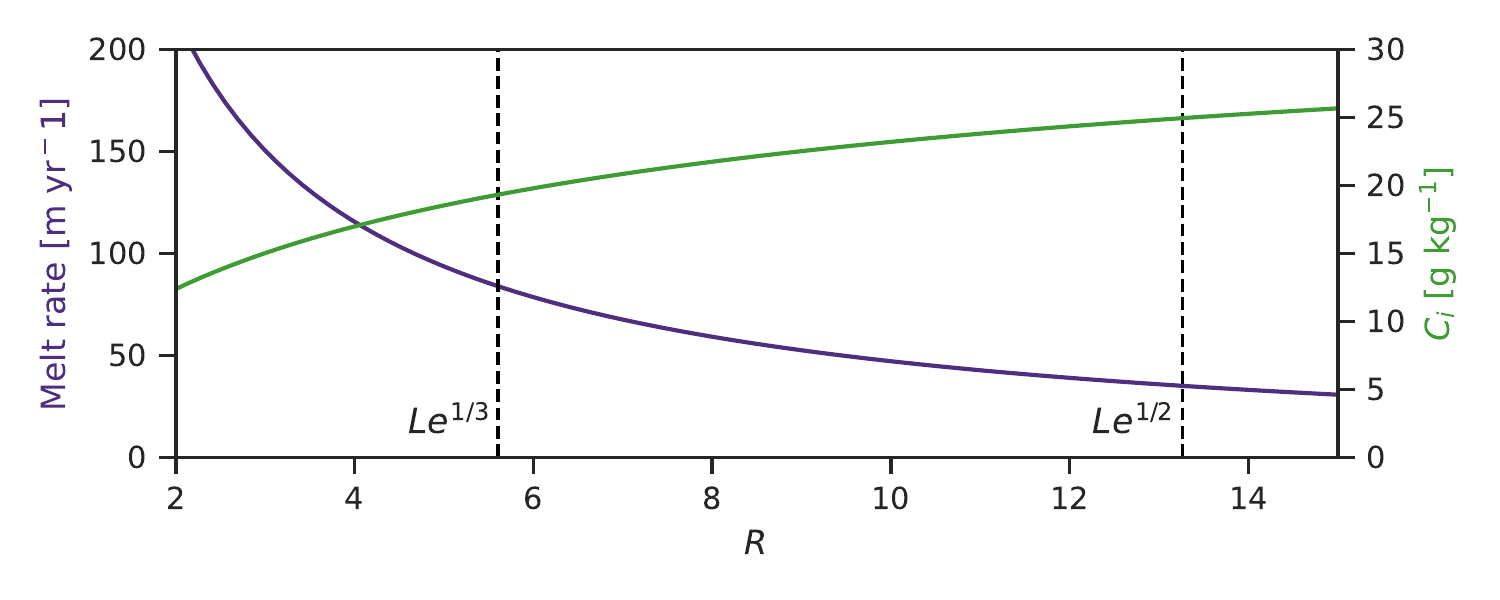}
    \caption{
            Dependence of melt rate and interface salinity $C_i$ on flux ratio $R$.
            The ambient ocean salinity is assumed to be fixed, so a lower $C_i$ leads to a stronger buoyancy source of fresher water at the ice face, driving stronger convection and enhancing the melt rate.
            Vertical dashed lines mark the two values of the flux ratio assumed by \citet{jenkins_convection-driven_2011} ($Le^{1/3}$) and \citet{kerr_dissolution_2015} ($Le^{1/2}$).
            This result assumes (i) salt flux is determined by $Nu\approx 0.1 Ra^{1/3}$, (ii) ambient ocean temperature of \SI{1}{\celsius}, (iii) physical constants prescribed as in table \ref{tab:dimensional_quantities}.
        }
    \label{fig:melt_R_dep}
\end{figure}

Using the physical parameter values in table \ref{tab:dimensional_quantities}, we find that for an ambient ocean temperature of $T_\infty=\SI{1}{\celsius}$, the interfacial salinity $C_i$ varies significantly with the Nusselt number ratio $R$.
For $Le=175.8$, taking $R=Le^{1/3}=5.6$ gives an interface salinity of $C_i=\SI{19.3}{g.kg^{-1}}$, whereas following \citet{kerr_dissolution_2015} and taking $R=Le^{1/2}=13.25$ gives a result of $C_i=\SI{24.9}{g.kg^{-1}}$.
This in turn leads to an even greater effect on the melt rate.
As a crude estimate for the salt flux, we can take the estimate $Nu_C\approx 0.1 Ra^{1/3}$, although such a simple power-law description does not fully describe the vertical convection system \citep{howland_boundary_2022}.
Applying the values in table \ref{tab:dimensional_quantities} leads to melt rate predictions from this simple model that vary from $\mathcal{V}=\SI{35}{m.yr^{-1}}$ with $R=13.25$ up to $\mathcal{V}=\SI{84}{m.yr^{-1}}$ when using $R=5.6$.
A wider dependence of the melt rate (and interface salinity) on the flux ratio is shown in figure \ref{fig:melt_R_dep}.

This factor of more than two in ablation velocity highlights the sensitive nature of melt parameterisations to the physical assumptions underlying them.
More research is undoubtedly needed to couple numerical results and theory with experiments and observations.
In particular, the transition between convectively-driven and shear-driven flows, where these different flux ratios appear relevant, must be understood.
This has practical importance for the case in which steep ice faces are subject to horizontal flows in conjuction with the vertical convection of the meltwater - a case of mixed convection \citep{jackson_meltwater_2020}.
Although, from our results, the $Le^{1/3}$ flux ratio scaling appears relevant to both convective and sheared systems, the functional form of a melt parameterisation that applies universally remains uncertain \citep{mcconnochie_testing_2017}.
It will be useful to consider a variety of geometries in such process studies.
This work focused on the symmetric case of a vertical channel to obtain temporally converged statistics, but it is unclear how exactly the lateral confinement imposed by the walls may affect the boundary layers when compared to a growing wall plume \citep{van_reeuwijk_mixing_2019,ke_high_2021}.
In environmental scenarios, the ice surface is also rarely smooth, with distinctive scallop-like roughness seemingly ubiquitous on the underside of icebergs \citep{bushuk_ice_2019}.
A full understanding of the ice-ocean boundary layer will be incomplete without a physical description of this complex two-way coupling between the flow and the shape evolution of the solid phase.
Promising advances in recent work are already enhancing our understanding of the coupled morphodynamics of ice in the presence of shear flows and convecting melt \citep{couston_topography_2021,weady_anomalous_2022,wang_equilibrium_2021,ravichandran_combined_2022}.

\appendix
\section{Temperature variance budget}

\begin{figure}
    \includegraphics{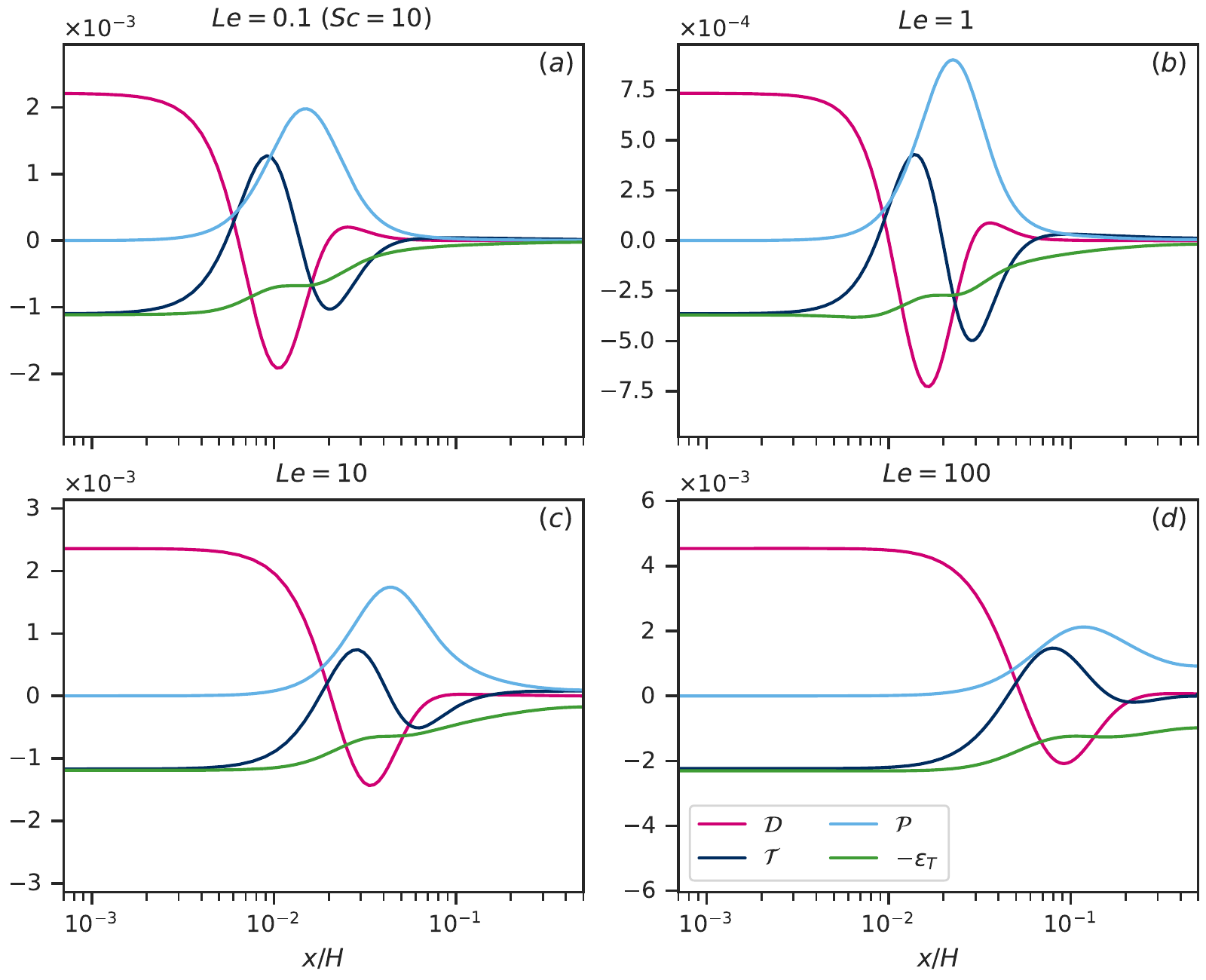}
    \caption{
        Wall-normal profiles of the scalar variance budget terms from \eqref{eq:var_budget} for three simulations:
        $(a)$ A10L01
        $(b,c)$ A100L10
        $(d)$ A100L100.
        In $(a,c,d)$ the temperature variance budget is presented.
        In $(b)$ the budget terms plotted are actually those of the concentration variance to provide insight into the dynamics at $Le=1$, in which case temperature statistics would evolve identically to the concentration statistics shown in panel $(b)$.
    }
    \label{fig:var_budget}
\end{figure}

As an extension to the heat flux analysis in section \ref{sec:results}, where we consider the budget terms for the mean temperature equation, we can investigate the terms contributing to the evolution of the temperature variance.
This informs us about the mechanisms driving, transporting and dissipating turbulent thermal fluctuations through the system.
We begin by decomposing the temperature field as before into a mean component and its fluctuation, where the mean is taken in the homogeneous directions $y$ and $z$:
\begin{equation}
    T(x,y,z,t) = \overline{T}(x,t) + T'(x,y,z,t) .
\end{equation}
By multiplying \eqref{eq:Tevo} by $T$ and decomposing the temperature field as above, we can derive the evolution equation for the temperature variance as
\begin{equation}
        \frac{\pd }{\pd t} \frac{\overline{{T'}^2}}{2} =
        \underbrace{\kappa \frac{\pd^2 }{\pd x^2} \frac{\overline{{T'}^2}}{2}}_{\mathcal{D}} \quad + \quad
        \underbrace{- \frac{\pd}{\pd x} \overline{u'\frac{{T'}^2}{2}}}_{\mathcal{T}} \quad + \quad
        \underbrace{- \overline{u'T'}\frac{\pd \overline{T}}{\pd x}}_{\mathcal{P}} \quad - \quad
        \underbrace{\kappa \overline{|\nabla T'|^2}}_{\varepsilon_T} ,
    \label{eq:var_budget}
\end{equation}
The budget terms on the right hand side can be interpreted respectively as the diffusion and transport of temperature variance, the production of temperature variance by the mean temperature gradient, and the dissipation of temperature variance.
Since the system reaches a statistically steady state, the budget terms must sum to zero at every wall-normal position.
We also note that the volume integrals of the transport and diffusion terms $\mathcal{T}$ and $\mathcal{D}$ must be zero, so globally there is a simple balance between production $\mathcal{P}$ and dissipation $\varepsilon_T$.

In figure \ref{fig:var_budget} we plot the various budget terms as a function of $x$ for a selection of simulations at various Lewis numbers.
Overall, the structure of the budgets appears very similar in all these cases.
The production is localised with a distinctive peak, but this is not balanced locally by dissipation.
Instead, there are also significant negative contributions from the diffusion and transport terms.
We can use figure \ref{fig:passive_comp}(b,d) to provide further interpretation for the location of the production peak.
In that figure, we observe a peak in the temperature (and concentration) variance that moves further from the wall as $Le$ increases.
This variance peak coincides with the minimum of the diffusion term in figure \ref{fig:var_budget}, which is slightly closer to the wall than the peak in variance production.
In the bulk, the various budget terms are small at low $Le$, but as $Le$ increases and the production peak moves further from the walls, the production and dissipation at the channel centre become more significant.
Despite the localised peak in variance production away from the wall, the peak value of its dissipation occurs at the wall as $x\rightarrow 0$.
Here the transport and dissipation terms are equal and opposite, balancing the contribution from diffusion in every simulation.
In this near-wall sublayer region, these quantities are roughly constant, suggesting that the rms temperature fluctuation scales linearly with distance from the wall.\\\



The data used to construct the figures in the paper is openly available at \citep{Howland_Data_supporting_Double-diffusive_2022}.

\begin{acknowledgments}
    We thank two excellent anonymous reviewers for improving the focus and clarity of the paper through their insightful and thoughtful comments.
    This project has received funding from the European Research Council (ERC) under the European Union's Horizon 2020 research and innovation programme (Grant agreement No. 804283).
    We acknowledge PRACE for awarding us access to MareNostrum at Barcelona Supercomputing Center (BSC), Spain (Project 2020235589).
    This work was also carried out on the Dutch national e-infrastructure with the support of SURF Cooperative.
\end{acknowledgments}
\newpage

\bibliography{DDVC_paper.bib}

\end{document}